# A FERMI SEA OF HEAVY ELECTRONS (A KONDO LATTICE) IS NEVER A FERMI LIQUID

ABSTRACT---

**I demonstrate a contradiction which arises if we assume that the Fermi surface in a heavy electron metal represents a finite jump in occupancy.**

------
The striking quantum phenomenon of heavy electron formation occurs in intermetallic compounds of the rare earth and actinide metals (usually Ce, Lu, and U, but occasionally others). The f-shell electrons, which are, at room temperature, to all intents localized spins, scattering a conventional sea of free metallic electrons, cross over at low temperature into mobile band electrons, albeit with very heavy effective masses, and change the Fermi surface radically in order to accommodate precisely the number of electrons—or, in the case of Lu, holes—which accounts for the number of spins.[1]

From a fundamental point of view the most surprising feature of this observation is that the dimensionality of the Hilbert space which we must use to describe the wave function has radically changed. N sites on which we may have a spin up or down have $2^N$ possible states; but if we can occupy each of these N sites with 0, 1 or 2 real electrons that amounts to $4^N$ possible states. (If there is orbital degeneracy that merely changes the arithmetic, not the enormous discrepancy in dimension.) It turns out that this is the crucial feature: enforcing the requisite constraints on the $4^N$ degrees of freedom causes a characteristic anomaly. The net effect will be to make the T=0 axis into a critical line, having a continuously variable exponent for some properties, but not to invalidate Luttinger's theorem and the existence of a Fermi surface.

It is essential to go into the physics of why Ce f electrons, for example, normally act as spins. The 5f shell is deeply localized within the atom, so that two f electrons on the same atom will interact strongly via their Coulomb repulsion, which is not effectively screened by outer-shell, metallic electrons. Although the f electrons can hybridize to an extent with the metal band electrons, the breadth of the f band caused by this will undoubtedly be small compared to the repulsion. There will be no more than one f electron per site, which may have up or down spin. This is often modeled as simply a spin (the "Kondo lattice" model), but one should not forget that in condensed matter situations spins are always electrons. One correct way to express this fact is to describe the spin as a projected electron, i e to project away the possibility of double occupancy of the f shell from a wave function written nominally in the full Hilbert space of four possible occupancies.

If the spins are dense, as we lower the temperature they will tend to order magnetically, and indeed the rare earth elements are almost all magnetic. But in a sufficiently dilute compound the magnetic interactions may not dominate, and the hybridization with the metallic electrons may lead to the formation of narrow f bands. In fact, naïve band calculations always predict f bands much less narrow than are observed, and that the mobilization of the f electrons should be much easier than it is; but spectacularly, wrong as such calculations are quantitatively, they predict the correct size, and often shape, of the Fermi surface which eventually appears at low temperature.

I said we must project out double occupancy; when there is one electron per atom this has the effect of requiring one spin per site. When the bands form, a few f-sites will empty because they are hybridized with the band electrons, and one can show that the chemical potential will be such that the energy of an empty site is not far from that of a singly-occupied one, and should not be excluded by projection; so in fact we must project down to a space of 3 states per f-site, not 2. This implies that the ground state wave function may be written

$$\Psi(r_1,\sigma_1;r_2,\sigma_2;.....) = P_G\Phi(r_1,\sigma_1;.....) \; with$$

$$P_G = \prod_i (1 - n^f_{i\uparrow} n^f_{i\downarrow}) \qquad [1]$$

Here $\Phi$ is a general function in the full Hilbert space of all electrons' coordinates and spins, and the "Gutzwiller" projector $P_G$ projects out all doubly-occupied atomic states. A great deal of misinformation is in the literature about this projection process, and we should emphasize two points: First, that it can be derived as a non-singular canonical transformation; and second, any perturbative admixture between the two subspaces of states can only increase the separation to the "upper Hubbard band" of doubly-occupied states.

The theory of the "freezing-out" process for the spins, (sometimes called "Kondoization") has been the subject of a large literature, but that is not our concern here. The introduction of the Gutzwiller projection referred to above into the theory is due to Rice and Ueda[2]; and the most reliable quantitative account of the process of forming heavy electron bands is given by the dynamical mean field theory of Kotliar and Georges[3]. But what we are here concerned with is the end product at absolute zero: the ground state and low elementary excitations.

Most of the heavy-electron materials have either superconducting or magnetically ordered ground states. (Often called "animal" and "mineral") A few, however, persist as supposed Fermi liquids to absolute zero; while also a number have competing and incompatible magnetic and superconducting orderings and therefore at the critical point between them have, again, no order at all (vegetable). While the effects we will discuss persist in altered form into the other phases, it simplifies matters to study the case of no ordering. Our method of proof is to assume there is a Fermi liquid ground state and show that leads to a contradiction.

A Fermi liquid may be treated as a non-interacting system with quasiparticles described by Fermion operators $c^*_{i,\sigma}$, $c_{i,\sigma}$ in site representation, and correspondingly

$$c_{k,\sigma} = \frac{1}{\sqrt{N}} \sum_i e^{-ik\cdot r_i} c_{i,\sigma} \quad [2]$$

etc in momentum representation. The Green's function is defined in space-time as (omitting spin indices as irrelevant)

$$G_{ij}(t;t') = -i\langle 0 | T(c_i(t) c^*_j(t')) | 0 \rangle \quad [3]$$

|0> is the ground state. (Most of what follows refers to the site-diagonal Green's function $G_{ii}$, but the analytic structure is general.) G may also be represented in frequency-momentum space by its Fourier transform $G(k,\omega)$. G has a well-known representation in terms of the densities $A(\omega)$ and $B(\omega)$ of excited states accessed by adding or removing one Fermion:

$$G_{ij}(\omega) = \int_0^\infty [\frac{A_{ij}(E)}{\omega - E + i\delta} + \frac{B_{ij}(E)}{\omega + E - i\delta}] dE \quad [4]$$

(we set the Fermi level at $\omega=0$). Here A and B are defined by

$$A_{ij}(E) = (2\pi)^3 \sum_s (0|c_i|s)(s|c^*_j|0) \delta(E - \varepsilon_s)$$

$$B_{ij}(E) = (2\pi)^3 \sum_s (0|c^*_i|s')(s'|c_j|0) \delta(\varepsilon_{s'} + E) \quad [5]$$

That is, they are the densities of, respectively, electron and hole eigenstates s and s' at energy E accessed by applying an electron or hole creation operator to the ground state. From [4] and [5] we can derive the real and imaginary parts of G and a dispersion relation between them:

$$\mathrm{Im}\, G_{ij}(\omega) = \{-\pi A, \omega > 0; \pi B, \omega < 0\} \text{ and}$$

$$\mathrm{Re}\, G_{ij}(\omega) = \frac{P}{\pi} \int_{-\infty}^{\infty} \frac{\mathrm{Im}\, G(\omega')\,\mathrm{sgn}(\omega')}{\omega'-\omega} d\omega' \qquad [6]$$

[6] is not the conventional Hilbert transform because the singularities of G cross the real axis at $\omega=0$; but correspondingly A and B change sign at $\omega=0$ and the integrand is normally non-singular. This absence of singularity is only assured if $A(0) = B(0)$, a requirement which is taken for granted in the standard texts,[4] and is obviously true in the Fermi liquid, where the hole quasiparticle is simply the negative of the electron at the same momentum. What will be shown here is that this is not true for projected electrons, i e spins; and that therefore [4] is logarithmically singular: the assumption that there is a Fermi liquid is mathematically inconsistent.

Let us consider the fermion creation and destruction operators for an electron in the f shell on a particular site i: $c^*_{i,\sigma}$ and $c_{i,\sigma}$. The presumed band state $|k,\sigma\rangle$ will be created or destroyed by a linear combination of these operators as in equation [2]. The matrix elements which enter into the definitions [5] of A and B are those of these operators, acting on the *projected* wave-function $\Psi$. The reason for a difference is almost obvious: a hole can always be created without violating the constraint, but an electron often encounters an already-occupied site.

Formally,
$$c^*_{i,\sigma} \Psi = (1 - P_G) c^* P_G \Phi + P_G c^* P_G \Phi$$

and the first term is finite but creates states in the upper, split-off Hilbert space which cannot contribute to A near ω=0; these states have $\varepsilon_s$ near or above the large repulsion U.  On the other hand, $c_{i,\sigma}\Psi = P_G c_{i,\sigma} P_G \Phi$ , since it cannot create double occupancy

Note that

$$P_G c^* P_G = (P_G c P_G)^*;$$ [7]

These operators are conjugates of each other and would have equal A and B densities; c* and c do not.

To illustrate the difference let us make the "Gutzwiller approximation" of assuming that Φ, like a simple Slater determinant of Bloch waves, has no correlation of site occupancies for opposite spins. (If there is a "correlation hole" the numbers are modified but the structure is the same.  The basic point is that site occupancy probabilities aren't singularly dependent on momentum.)  Let us define a parameter x which is the number of holes relative to the case of 1 f electron per site. Every state s is equally accessed by the local Fermion operator, and there are (1-x)/(1+x) fewer full states below $E_F$ than empty ones above it, in the state Φ.

In the state Ψ, on the other hand, there are only x empty ones to (1-x)/2 full ones, so that the weight of the terms of equation [5] in A is reduced relative to those in B by the ratio

$$A/B = g = \frac{x}{(1-x)/2} \times \frac{1-x}{1+x} = 2x/(1+x)$$ [8]

There will thus be a logarithmic singularity in the real part of G, according to equation [6], proportional to 1-g=(1-x)/(1+x) (times the renormalization constant Z which gives the height of the Fermi discontinuity, in case there are many-body corrections.)

This singularity is obviously not allowable.  If A(ω=0) is related to B by a constant ratio, as in [8], the only way in which the singularity can be avoided is if A=B=0.  The conjecture I have arrived at through more complicated and less rigorous reasoning[5] is that the real and imaginary parts of G have primarily a power law dependence on ω for small values, with a small but finite positive exponent.  There is strong experimental evidence for such a power law singularity in the case of the cuprates (see ref [5]) but the direct measurement of the Green's function has not yet been carried out for heavy electron materials.  Indirect evidence in the form of transport anomalies does exist.[6]

In the heavy electron case, there has been much discussion in the literature of possible "quantum critical point" effects.  I feel that one must first clear up the effects of the above universal critical line before speculating about QCP's.  A question which will occur to many readers is  why the transport properties often resemble those of a Fermi liquid except near QCP's.  As in the cuprates, this has to do with details  of the transport theory; our conjecture is that there is a "hidden Fermi liquid"[7] which can constitute a "bottleneck" for some relaxation processes in some circumstances, and make the latter look like those of a Fermi liquid.

---

[1] A review is N Grewe and F Steglich, Handbook on Phys and Chem of Rare Earths, v 14, eds K A Gschneider and L Eyring, p343 (Elsevier, Amsterdam, 1991)

[2] T M Rice and K Ueda, 1986, Phys Rev B34, 6420 (1986)

[3] A Georges, G Kotliar, W Krauth, M Rosenberg, Revs Mod Phys 68, 13 (1996)

[4] A A Abrikosov, L P Gor'kov, I Dzialoshinsky, "Methods of Quantum Field Theory in Statistical Physics", Prentice-Hall NJ, 1963.  The equations used here are adapted from this text.

[5] P W Anderson, Nature Physics 2, 360 (2006); see also P A Casey et al, Nature Physics 4,  (2008)

[6] P Gegenwart,  Qimiao Si, F Steglich, Nature Physics 4, 186 (2008).

[7] P W Anderson, cond-mat  0709.10339; submitted to Phys Rev B